\UseRawInputEncoding
\documentclass[%
 reprint,
 amsmath,amssymb,
 aps,
prx,
]{revtex4-2}

\usepackage{physics}
\usepackage{siunitx}
\usepackage{graphicx}
\usepackage{dcolumn}
\usepackage{bm}

\usepackage{bm}

\usepackage[normalem]{ulem}

\def\add#1{{\textcolor{black}{#1}}}    
\def\del#1{{}}  

\begin{document}

\preprint{APS/123-QED}

\title{Advancing flight physics through natural adaptation and animal learning}

\author{Ariane Gayout}
 \email{a.m.m.gayout@rug.nl}

\affiliation{%
Energy and Sustainability Research Institute Groningen, Faculty of Science and Engineering, \\
University of Groningen, 9747 AG Groningen, The Netherlands
}%
\altaffiliation[Also at ]{%
 Groningen Institute for Evolutionary Life Sciences, University of Groningen, The Netherlands
}

\date{\today}

\begin{abstract}
Fluid dynamics, and flight in particular, is a domain where organisms challenge our understanding of its physics. Integrating the current knowledge of animal flight, we propose to revisit the use of live animals to study physical phenomena. After a short description of the physics of flight, we examine the broad literature on animal flight focusing on studies of living animals. We start out reviewing the diverse animal species studied so far and then focus on the experimental techniques used to study them quantitatively. Our network analysis reveals how the three clades of animals performing powered flight - insects, birds and bats - are studied using substantially different combinations of measurement techniques. We then combine these insights with a new paradigm for increasing our physical understanding of flight. This paradigm relies on the concept of Animal Learning, where animals are used as probes to study fluid phenomena and variables involved in flight, harnessing their natural adaptability.
\end{abstract}

\maketitle


\section{\label{sec:nature}Learning about Physics from Nature}

\del{Physics, from its Latin origin \textit{physica}, is the study of nature} \add{Going back to the Latin origin of physics, \textit{physica}, nature constituted its initial object of study}. Yet nowadays, observing nature itself to study physics has become secondary to building complex systems in laboratory conditions or numerical simulations. This provides support to theoretical hypotheses and technological development, with little to no equivalent of such system in any natural environment.
 In contrast, fluid dynamics is one example of the few physics domains that still retains natural observations at the core of its understanding.
 
\begin{figure}[ht!]
    \centering
    \includegraphics[width=0.95\linewidth]{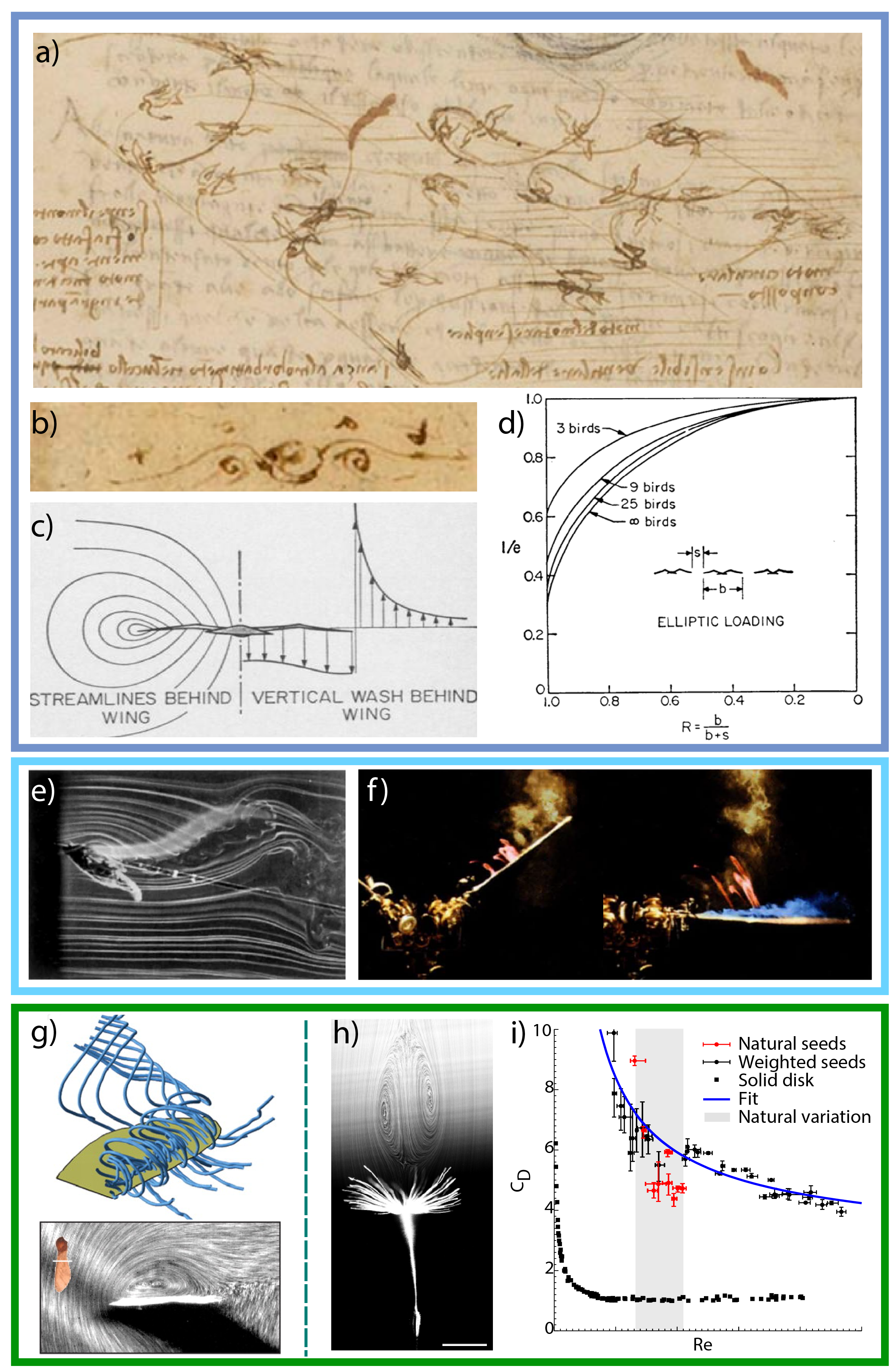}
        \caption{Overview of aerodynamic mechanisms brought to light through observations of natural phenomena. Looking at how birds circle in the atmosphere, Leonardo da Vinci first described thermals (a). Imagining how the weight of the bird must be balanced by the air, he proposed the wings must generate a pressure difference  (b) [codexatlanticus.it: a) p.845, 1505; b) p.1098, 1514]. Observing how birds prefer flying in V-formation stimulated analyzing how their wakes interact with the wing, it was found that lift induced drag is mitigated by adjacent wingtip vortices (c-d) \cite{Lissaman1970}. By visualizing the flow around an insect wing (e), and using a dynamically scaled robotic model (f) \cite{Ellington1996}, an unexpected leading-edge vortex was discovered to increase lift beyond values according to aircraft theory. Subsequently it was discovered across organisms including rotating seeds (g) \cite{Lentink2009}. Finally, the microstructure of a dandelion seed enhance the aerodynamic drag in an unexpected way by stabilizing a ring vortex in its near-wake that increases pressure drag (h-i) \cite{Cummins2018}. All images are reproduced with permission.}
    \label{fig:lit}
\end{figure}

The value of observations in fluid dynamics is signified by the seminal work of Leonardo da Vinci, which \add{founded and} still inspires \add{fluid dynamics, and especially} flight, research (Fig.\ref{fig:lit}.a-b). In particular, tracking gliding birds, he first described atmospheric thermals (Fig.\ref{fig:lit}.a) and later suggested that vortex generation and pressure differences may be responsible for lift production in birds (Fig.\ref{fig:lit}.b).
Birds have continued to inspire our physical understanding of wake interactions in an effort to understand V-formation flight by Lissaman and Shollenberger in 1970 (Fig.\ref{fig:lit}.c-d) \cite{Lissaman1970}. By flying together, birds save on drag and thus on energy expenditure (Fig.\ref{fig:lit}.d), thanks to the downwash interaction induced by their wingtips (Fig.\ref{fig:lit}.c).

While birds are often presented as the main contributor to our understanding of flight, insects have revolutionized the understanding of vortex-induced lift generation. This phenomenon, brought to light by Ellington \textit{et al.} in 1996 (Fig.\ref{fig:lit}.e-f) \cite{Ellington1996}, resolved a physical interrogation raised in 1934 by A. Magnan, that bumblebees were not suppose to fly and were defying physical laws \cite{Magnan1934}. Lift in insects is predominantly generated by the leading-edge vortex that remains stable by the revolving wing when flapping around its hinge. 
Such mechanism has later been confirmed to be widely used in natural flight \cite{Chin2016}, even in plants like rotating seeds (Fig.\ref{fig:lit}.g) \cite{Lentink2009}.


Other aerodynamic phenomena have been derived from plants, like the effect of microstructure stabilizing large vortex structures at low Reynolds numbers, in dandelion seeds (Fig.\ref{fig:lit}.h-i) \cite{Cummins2018}. Microstructures \add{have also been reported to play a role in silencing owl flight \cite{Sagar2017,Wagner2017,Jaworski2020}} and enable the flight of millimetric insects \cite{Farisenkov2022}.
\\

These key examples show how studying organisms has deepened our understanding of the physics of flight beyond engineering. To build off this, we first review the current knowledge of animal flight physics (\ref{sec:flightphysics}). We then analyze the context of live animal flight experimentation (\ref{sec:landscape}). We follow on which experimental environment (\ref{subsubsec:conditions}) and techniques have been used to study the few animal species considered till date (\ref{subsubsec:tech}) and find surprising differences between clades. Considering the large number of species and behaviors that remain to be studied using the same methods, we identify significant opportunities for discovering new flight physics (\ref{subsec:future}). Finally we present a new approach for effectively perusing the vast parameter space, animal learning, by harnessing animal intelligence and natural adaptation as experimental probes (\ref{sec:AL}).

\vspace{-0.4cm}
\section{\label{sec:flightphysics}The Physics of animal flight}

\del{In the 19th century, after Euler, Navier and Stokes developed the fluid equations of motion, Helmholtz proposed the first scaling laws for \add{animal} flight \cite{Runge1893}. Modern data-driven scaling laws for flight, integrating wing loading in particular, have then been developed by Tennekes in 1996 and Lindhe Norberg in 2006.}
\add{From the founding work of da Vinci, fluid dynamics was further developed by Newton, d'Alembert and many others before Euler, Navier and Stokes formalized the equations for fluid motion. These advancements led to the establishment of the first scaling laws to animal flight by Helmholtz in the late 19th century \cite{Runge1893}. As aviation soared in the 20th century with Lilienthal and the Wright brothers, aerodynamic research saw the rise of parametric studies, driven by the National Advisory Committee for Aeronautics in the United States to which we owe the NACA profiles and the Kaiser Wilhelm Institute for Flow in Germany with Prandtl and von K\'arm\'an \cite{Winter1936,Flachsbart1932,Hoerner1965,wright1953papers,Foppl1911,Eiffel1910}. These parametric studies initiated data-informed modeling that materialized in animal flight as data-driven scaling laws, integrating wing loading in particular \cite{Tennekes1996,LindheNorberg2006}.}
In their current form, these scaling laws have been proven valuable for understanding airplane design and animal flight, \add{to which} further development can make them valuable for flight physics \cite{Sethna2022}.

A particularity of animal flight as a research domain is indeed its intrinsic multidisciplinarity, ranging from physics and biology to engineering. As such, it is not surprising that models developed for animal flight were proposed by biologists, like Pennycuick or Weis-Fogh \cite{pennycuick_book,Weis-Fogh1973}, while others emanate from fluid dynamicists \cite{Wang2005,Liu2024}.
\\

 \del{Free flight in the free atmosphere is primarily governed by two pressure-dominated} \add{In the context of free flight, animals are primarily influenced by two aerodynamic} forces: drag and lift. \add{As flying animals size from a few meters down to hundreds of micrometers, the nature of these forces is highly dependent on the animal size, more precisely on its Reynolds number $\mathrm{Re}=UL/\nu$, with $U$ the air velocity in flight, $L$ the size of the animal, typically its wingspan, and $\nu$ the kinematic viscosity of air.}
\add{In particular, large gliding birds ($\mathrm{Re}\sim 10^5$) often generate lift thanks to their wing cambers like planes. }

\add{This camber-induced lift is yet often coupled with a strong leading-edge vortex, that stabilizes a depression zone along the wing enhancing lift. Similar leading-edge vortex is found in most animal flights, even when camber-induced lift is absent, as we mentioned earlier \cite{Chin2016}. Such leading-edge-vortex-induced lift is at the core of lift production in flapping flight, as flapping induces a cycle of positive and negative lift production alternating between the downstroke and upstroke. This cycle can lead to extreme lift fluctuations, especially for insects like butterflies \cite{Langley2014}.}
\add{Another vortex-induced lift mechanism is found in gliding vertebrates of small aspect ratio, where trailing vortices developed from the wingtips enhance lift at high angles of attack, similarly to delta wings \cite{Thorington1998,Lau2023}.}

\add{Vortex-generated lift, be it from the leading-edge or the wingtips, is coupled with a pressure-induced drag due to the angle of attack of the wing. Such drag may however harnessed as a force along the wing normal for lift production, especially at landing \cite{Chin2019a}.}
\add{For tiny insects however, vortex-induced aerodynamic forces are less relevant due to their low Reynolds number $\mathrm{Re<40}$. On the contrary, friction has recently been observed to be determining in the lift and drag production with high shear stresses \cite{Liu2024,Farisenkov2022}.
}
\\

\add{In addition to the mechanisms from which lift and drag originate, which force predominates also plays a role on the flight physics. In particular,} an animal can be parachuting, when drag overcomes lift, or flying, when lift prevails. A simple separation between these two regimes is the descent angle, or glide angle. Parachuting then corresponds to angles above \SI{45}{\degree} \cite{Oliver1951}.
While parachuting includes a wider diversity of animals \cite{Dudley2011}, we will focus on lift-dominated flight, meaning powered and gliding flights, for the rest of this article.

\begin{figure}
    \centering
    \includegraphics[width=0.95\linewidth]{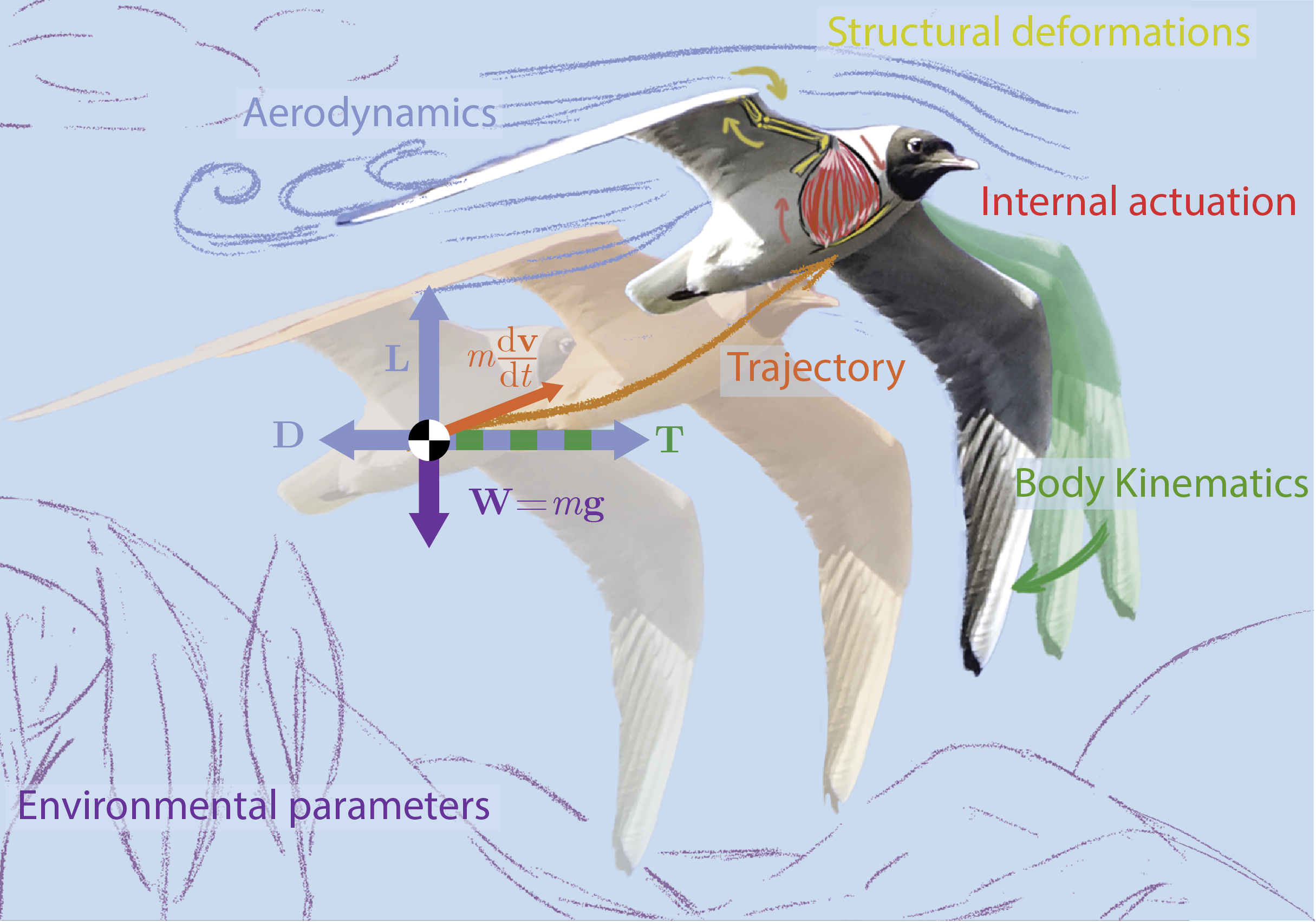}
    \caption{Animal flight results from a force balance and involves physical variables including air velocity and pressure, observables (measurable variables) such as kinematics and constraints linked to the environment, like gravity and the airscape. Illustration for a black-headed gull \textit{Chroicocephalus ridibundus}.}
    \label{fig:flight}
\end{figure}

Putting flight into an equation is often as difficult as interpreting its scaling laws, due to the multi-level structure of interactions at play. The equations of flight may be, for instance, derived from Newton's second law of motion:
\begin{equation}\label{eq:newton2}
    m\dv{\textbf{v}}{t}=\sum \textbf{F}_{ext}
\end{equation}

where $m$ is the mass of the animal, $\textbf{v}$ its velocity, $\textbf{F}_{ext}$ any external forces acting on the animal. 
From a mechanical point of view, $\sum \textbf{F}_{ext}$ includes in particular aerodynamic forces, lift ($\textbf{L}$), thrust ($\textbf{T}$) and drag ($\textbf{D}$), and gravity $\textbf{W} = m\textbf{g}$, as shown in Fig.\ref{fig:flight}. 
While lift, thrust and drag belong to aerodynamics observables (blue), weight is a known environmental constraint \del{, as gravity is an environmental parameter} (purple). Their balance then results in a specific trajectory (orange).

While this force balance helps to understand large-scale trajectories, it reduces the animal to its center of mass. Yet, animals are morphable solid objects with variable moment of inertia $I$. Hence, torque is also important to consider for stability. Torques add to forces the point of application, meaning that any morphing may affect not only the moment of inertia $I$ but also $\textbf{r}$, the position vector in the momentum equation:
\begin{equation}\label{eq:momentum}
    \dv{I\omega}{t}=\sum \textbf{r} \times \textbf{F}_{ext}
\end{equation}

An exhaustive model for flight should thus incorporate shape and inertia variations. These can be separated more particularly into body kinematics (green) and structural deformations (ocher). Body kinematics encompass flapping, dynamically morphing wings \cite{Pennycuick1968,Cheney2021,Harvey2022} or feet placements \cite{Combes2009}. Structural deformations relate closer to continuum mechanical models and consist of more subtle passive and active mechanisms, that can be observed on bat wing membranes \cite{Adams2012,Cheney2022} or bird wings at high angle of attack \cite{Pennycuick1982a,Carruthers2010}. In particular, while body kinematics may contribute largely to inertial changes, structural deformations typically primarily modify the aerodynamic characteristics of the animal itself \cite{Meresman2020,Cheney2022}. Body kinematics also impact drastically the aerodynamic forces and torques. They can modify the shape of the animal in the air or add temporal dynamics, thus increasing unsteadiness in the system.

Aerodynamic forces are indeed particularly sensitive to shape changes as variations of the aspect ratio of the wing can induce a shift from a \del{camber-induced lift to a vortex-generated lift predominance} \add{lift predominately generated by the leading edge vortex to one by the wingtip trailing vortices}. This comes from complex fluid-structure interactions, governed by Navier-Stokes equations, at play at the wing surface, which are yet solely quite understood for the leading-edge vortex \cite{Chin2016,Liu2024}.


On top of these animal components that influence aerodynamics, environmental parameters are of prime importance to truly appreciate the complexity of flight. \add{We consider here to be an environmental parameter, any variable that may play a role in the definition of the forces and is external to the animal and set by the conditions of experiment. In particular, gravity can be seen as an environmental parameters with even some experiments performed under microgravity conditions \cite{Oosterveld1975}.} Among these parameters, flow speed \cite{Hedenstrom2016,Sapir2014,Pennycuick1968a} and air density \cite{Altshuler2003,Dudley1995,Dillon2014} are commonly varied but turbulence \cite{Nasir2019,Crall2017} has recently been ``rediscovered" for its notable impact on insects and birds. Such rediscovery emanates in particular from biological considerations on human influence on its environment and bees struggling to adjust to higher turbulence levels \cite{Dargas2016,Hejazi2022,Ortega-Jimenez2018}, as well as engineering considerations for Unmanned Air Vehicles (UAV) bio-inspired control strategies \cite{Watkins2010}.
We can also add obstacles and maneuvering to these environmental parameters as they perturb the flow and the animal kinematics based on ``decision" from its end.
This can translate into choices in trajectories or flapping kinematics, which are two of the main observables in flight experiments with live animals as we will see in the following.

To complete the picture of the physics involved in flight, energy has to be considered as well. In particular, many studies investigate flight efficiency of animals, \textit{e.g.} through work loops \cite{Bahlman2020,Deetjen2024}, towards a better understanding of evolution and energy-saving strategies for airborne transportation like planes or UAV \cite{Nan2018}. The question on the energy consumption and internal actuation of the bird (red in Fig. \ref{fig:flight}) will be left out from this review as we focus more on the aerodynamics of flight.

With particular attention to the four observables, Trajectory, Aerodynamics, Structural deformations, Kinematics (TASK), we will see in the following that each can be measured through several experimental techniques, subject to yet additional implementation conditions compared to more standard physical studies, due to the presence of live animals.


\section{\label{sec:landscape}Current landscape of live animal flight study}


\begin{figure*}
    \centering
    \includegraphics[width=0.9\linewidth]{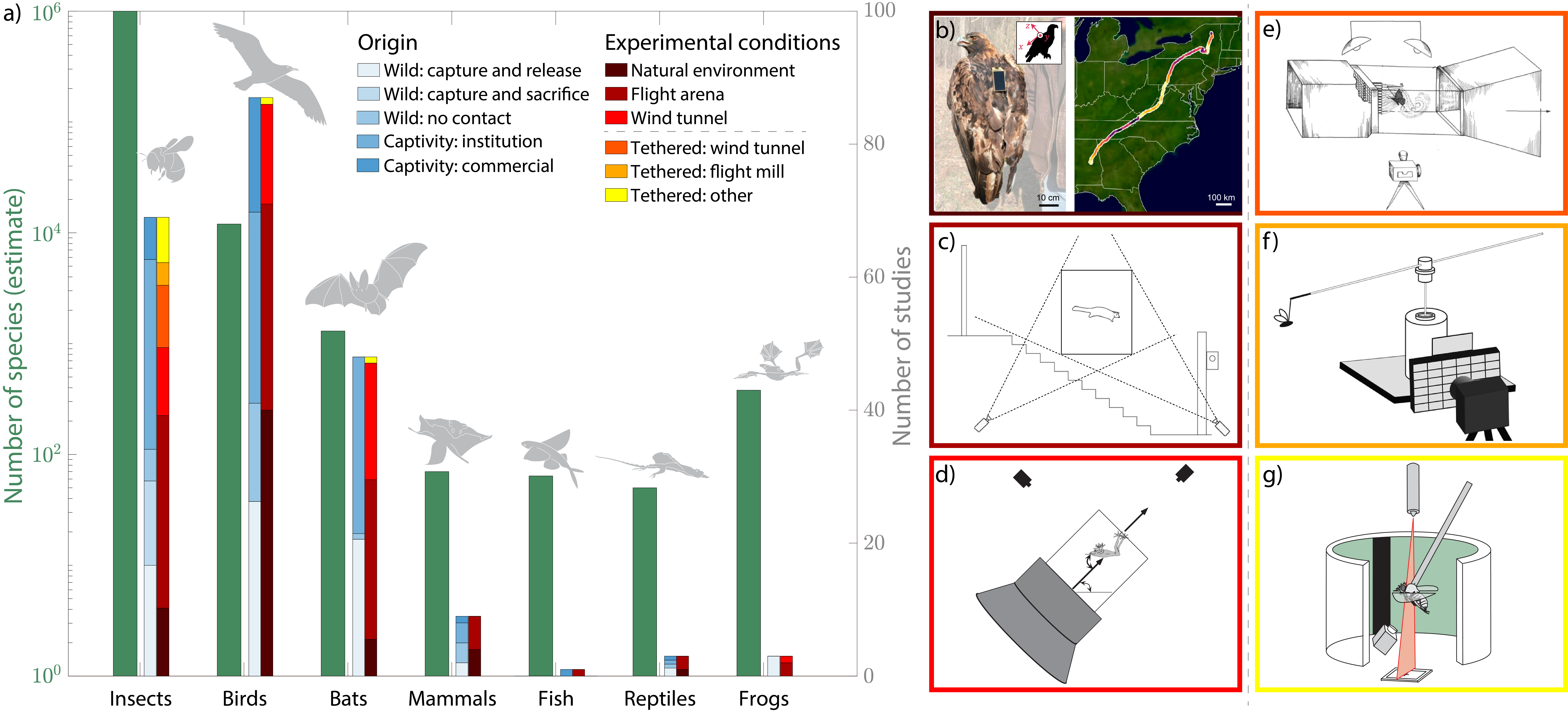}
    \caption{a) Discrepancy between flying animal diversity and the number of flight experiments conducted on each species group \textit{in vivo}. The number of species is given in green for each animal group on the left. The corresponding right bar indicates the origin of the animals used in blue shades (left) and the experimental conditions in which they were studied in red to yellow (right). Examples of experimental conditions are presented on the right (b-g): field conditions (b) \cite{Laurent2021}, lab flight arena (c) \cite{Bishop2006}, animal flight wind tunnel (d) \cite{McCAY2001}, and tethering of animals in wind tunnels (e) \cite{Brodsky1991}, flight mills (f) \cite{Hajati2023}, and flight simulators (g) \cite{Hesselberg2009}. All images are reproduced with permission.}
    \label{fig:species_and_origin}
\end{figure*}


Before going deeper into the details of the experiments, it is important to know more about what constitutes an animal flyer and how the current research has covered live (\textit{in vivo}) animal flight in experiments until now.
\\

Firstly, animals have evolved at least four times into powered flyers and many more times into gliders. 
While pterosaures are an extinct clade of powered flyers, insects, birds and bats form the three extant classes of such flyers. In fact, most animals are flyers. Flying insects are estimated to about a million species, segregated into at least 30 orders. Though not as numerous as insects, around 11 000 species of birds have been currently identified. Bats, with 1200 species, constitutes the second largest order among mammals, totaling about $20\%$ of all mammalian species.
In addition to that, there are about 70 species of gliding mammals: rodents and marsupials. Gliding extends also to fish with 64 species known for gliding behaviors, reptiles with 40 species and amphibians with about 200 flying frogs. However, flying frogs are mostly parachuting in the sense we define earlier and we will not dig too deep in this order.

To exemplify this diversity, Fig. \ref{fig:species_and_origin} presents the number of flying species (green, left axis) in each clade in logarithmic scale and compares it with the number of studies on their flight using live animals (in linear scale, right axis). 
In particular, live animal experiments studying flight hardly cover even a scratch of the biodiversity of animal flyers. While many studies have been conducted on insects, the sheer diversity of this order makes it a widely unknown territory. On the contrary, birds appear to be more represented in research on flight. This is a somehow biased result from our selection of live animal experiments. Insect flight studies often encompass multiple techniques compared to bird studies as we will see in the following and a large portion of our understanding of insect flight comes from scaled-up experiments \cite{Dickinson1999,Jardin2015} and CFD simulations \cite{Kolomenskiy2011,Engels2016,Kolomenskiy2016}, which are mostly not reported here when they are not connected to live animal experiments.
Moreover, vertebrate training and husbandry is time-consuming and the lifespan of these animals extends to multiple years to decades. Thus, birds and bats may be reused in several experiments over the years and fewer species may have been studied than for insects based on these aspects.

In particular, out of the about 250 articles reporting on animal flight experiments presented here \footnote{Please note that studies using live flying animals but not studying flight but decision controls, behaviors or biochemistry are not considered in this topical review. While we aim for this study to be as complete as possible, some articles may have been left out by inadvertence and we hope that the statistics we present here are as converged as they would be with all possible literature on the subject.}, only around 350 species are investigated at least once, with few extensively researched. These few species, like rock pigeon \textit{Columba livia} \cite{Corning1998,Dial1992,Pennycuick1968,Warrick1998,Pennycuick1968a,Biewener1998,Spedding1984,Ros2011,Lempidakis2022,Adams1999,MIZUGUCHI2017,Sankey2019,Taylor2019,Taylor2017,Sankey2019a,Gessaman1991,Krishnan2022,Williams2015}, bumblebee \textit{Bombus impatiens} \cite{Crall2017,Ravi2013,Mistick2016,Linander2016,Ravi2016,Buchwald2010}, zebrafinch \textit{Taeniopygia guttata} \cite{Tobalske1999,Labocha2015,Bahlman2020,Hambly2004,Hambly2002,Wang2018,Lapsansky2019,Nudds2003,Eckmeier2008,Tobalske2009,Tobalske2005,Crandell2015} or Pallas's long-tongued bat \textit{Glossophaga soricina} \cite{Muijres2011,Winter1998,Winter1999,Wolf2010,Hedenstrom2007}, can be considered animal models for the study of flight. 
As one of the difficulties of live animal flight experiment is the provision and husbandry of the animals, well-studied species like pigeons are often used thanks to existing breeding colonies in laboratories or wide commercial availability.

To better grasp this difficulty, the origin of the animals used in experiments is represented in blue shades (right axis) in Fig. \ref{fig:species_and_origin}. In particular, for insects, a majority ($52\%$) has been raised in captivity ($45\%$ for experimental purposes in animal facilities) but about a third of the insects caught in the wild were killed after the experiments for morphological parameter assessments. For birds, $49\%$ also come from captivity but two thirds of these birds only have been bred in universities and affiliated facilities while the last third has been commercially purchased. A significant portion of birds ($33\%$) has been captured from the wild and released after the experiment. Lastly the remaining $18\%$ has been studied without any contact, through radar or visual tracking, which we will detail in the next section. Bats are in majority ($56\%$) studied from captivity-reared colonies at animal facilities of universities and the rest has been captured and released from the wild. Note that all bat species present in Europe, Middle East and Northern Africa are protected by law, and 11 species in the US, which explains the absence of commercial availability of bats for experiments. On average, even gliding mammals and lizards present this pattern of half-wilderness and half-captivity, with few commercially available species. However, the reduced number of studies on these animals makes it difficult for these estimations to be statistically relevant.

In addition to the choice of the species and the origin of the animals, training may or may not be implemented for the experiment to be a success. Yet, considering animal welfare increasing awareness, training has become over the years standard practice on vertebrate flight experiments, initiated for birds by Pennycuick in 1968 \cite{Pennycuick1968a}. More recently, even insects have been trained for flight experiments, like bumblebees \cite{Mistick2016} or moths \cite{Ortega-Jimenez2013}. Training protocols are commonly described in the articles themselves or as supplementary materials. Complementary to these protocols, an overview of positive reinforcement training perspectives for birds can be found in Baker \textit{et al.} 2019 \cite{Baker2019}. 
While it may seem benign to train an animal for flight experiments as it is one of its natural behavior, some experimental techniques described hereafter require thorough training to be conducted, due to the non-natural environment in which the experiment is performed. The presence of the wind tunnel and the noise of the fans may alone be a source of stress to the animal, which would then not respond as desired in the study and may not fly at all. This could explain partly why some animals, like mammalian gliders, have never been flown in wind tunnels, and only in flight arena or in their natural habitat. The other reasons for such lack of experiments lie for instance in the protection status of some species, necessitating either special permits or collaboration with zoos. Gliding in wind tunnel for mammals and lizards is also difficult due to how they perform their glides, and requires tilting wind tunnels, to much higher degrees than used for bats and birds.


\section{\label{sec:design}Experimental designs for aerodynamic studies of animal flight}

\subsection{Experimental conditions}\label{subsubsec:conditions}
This last point brings us to discussing the environmental conditions in which the study is performed, as shown in Fig. \ref{fig:species_and_origin} in red to yellow shades (right axis). Flight experiments can be separated between free flight and tethered flight, depending on whether or not the animal is restrained in its motion. From an ethical point of view, tethered experiments induce discomfort to animals, which prevents the use of tether in vertebrate experimentation. Only two studies were found using a form of tethering in vertebrates, under anaesthesia for birds \cite{Bahlman2020} and using a leash for bats \cite{Panyutina2013}, compared to at least 24 studies involving tethered insects \cite{Lehmann1997,Lehmann1998,Wolf1993,Fu2022,Hesselberg2009,Kutsch2003,Fry2005,Thomas2004,Urca2021,Stavenga1993,Taylor2010,Taylor2003,Brodsky1991,Bomphrey2005,Ellington1996,Fuchiwaki2013,Mouritsen2002,Li2020,Hollick1940,Yokoyama2013,Jaroslawski2022,Hajati2023,Nedved2001,Henningsson2015}. Tethering in insect flight studies is found in several configurations serving different purposes. For instance, to study wake development, the insect will be tethered in a wind tunnel (Fig. \ref{fig:species_and_origin}.e). To investigate kinematics and force production, it can be placed on a flight mill (Fig. \ref{fig:species_and_origin}.f, see \cite{Attisano2015} for a detailed protocol). For maneuvering, flight simulators have been designed, based on optical flow integration by the insect in flight (Fig. \ref{fig:species_and_origin}.g).
Some of these 24 studies also present results from free flight experiments for comparison, as tethering often impacts on the insect response to the flow \cite{Fu2022,Kutsch2003,Fry2005,Urca2021,Thomas2004}.
\\

For free flight, different conditions are also available. For instance, studying animals in their natural environment can be seen as the most logical conditions when looking at biological aspects of flight. In particular, a large proportion of bird studies (45$\%$) were conducted in nature \cite{Flack2018,Bishop2015,Portugal2014,Taylor2017,Ling2018,Ling2019,Sankey2019a,Shelton2014,Sholtis2015,Hedenstrom2016,Hernandez-Pliego2015,Wehner2022,Voelkl2015,Taylor2019,Perinot2023,Sankey2019,DeMargerie2015,Lapsansky2019,Pennycuick2001,MIZUGUCHI2017,Hedenstrom2017,Pennycuick1960,Pennycuick1982a,Adams1999,Williams2020,Carruthers2010,Gessaman1991,Weimerskirch2016,VanDoren2016,Schmaljohann2008,Malmiga2014,Pirotta2018,Sapir2011,Alerstam2007,Backman2001,Galtbalt2021,Hawkes2017,Hernandez-Pliego2017,Cabrera-Cruz2019,Laurent2021}, using for instance GPS loggers (Fig. \ref{fig:species_and_origin}.b) \cite{Laurent2021}. Yet with about only $10-15\%$ of bat flight experiments \cite{OFarrell1977,Sapir2014,Hochradel2022,Luo2021,Adams2012,McCracken2016} and insect flight studies \cite{Hejazi2022,Crall2017,Dargas2016,BRADY1995,Combes2009,Sugiura2010,Srygley2007,SkowronVolponi2018,Chai1990,Ruppell1989,Cant2005,Aralimarad2011}, it is hardly the main experimental condition used for investigating flight physics. Indeed, for most animal groups, the flight arena is the main used condition, representing 42$\%$ of the total literature reviewed in this article. This can be understood from the versatility of a flight arena. Its layout can be tuned to suit natural behaviors such as distant poles mimicking trees to stimulate flight in gliding animals (Fig. \ref{fig:species_and_origin}.c) \cite{Bishop2008a}. The limited available space for the animal to fly in also facilitates its observation from multiple cameras and thus the reconstruction of its trajectory and body kinematics. Some flight arenas are even set within natural environment for studying environment-induced morphological differences or several sympatric species \cite{LeRoy2021,Sun2016,Ingersoll2018}. Flight arenas thus constitute 50$\%$ of bat studies \cite{Adams2012,Norberg1976,OFarrell1977,Aldridge1986,Luo2021,Barchi2013,Kugler2016,Winter1998,Voigt1999,Voigt2011,Riskin2012,MacAyeal2011,Hase2021,Hughes1991,Winter1999,Elangovan2007,Hughes1993,Lancaster1997,Schutt1997,Iriarte-Diaz2008,LindheNorberg2006,Windes2020,Sterbing-DAngelo2011,Voigt2013,Falk2015,Boerma2019,Ingersoll2018} and more than 35$\%$ of bird \cite{Matyjasiak2018,Williams2015,Wang2018,Hambly2002,Hambly2004,Eckmeier2008,Nudds2003,Chang2013,Tobalske2004,Labocha2015,Skandalis2024,Chin2017,Chin2019a,Ros2011,Rayner1991,Spedding1984,Biewener1998,Ortega-Jimenez2018,KOKSHAYSKY1979,Warrick1998,Ward1999,Usherwood2003,Sun2016,Tobalske2000,Altshuler2003,KleinHeerenbrink2022,Corning1998,Dial1992,Cheney2020,Cheney2021,Deetjen2024,Ingersoll2018} and insect flight research \cite{Vance2013,Ortega-Jimenez2018,Nasir2019,Ristroph2010,Sinhuber2017,VanderVaart2019,Ha2015,Linander2016,Srygley2004,Grodnitsky1996,LeRoy2021,Tanaka2010,Kang2018,Sridhar2021,Bode-Oke2020,Fry2005,Su2020,Langley2014,Ennos1989,Hollick1940,Srygley1999,Srygley2007,Srygley1990,Meresman2020,Heinrich1970,Kutsch1999,Kutsch2003,Kutsch1993,Hesselberg2009,Fu2022,Fischer1999,Dillon2014,Dudley1995,Cox1995,Buchwald2010}.
\\

To further investigate aerodynamic characteristics of flight, wind tunnels have been introduced in many forms with an entire literature of specific designs for animal flight studies \cite{Pennycuick1997,Hedenstrom2017a,Quinn2017,Breuer2022}. An often shared characteristic of animal flight wind tunnels is the possibility of tilting the flow to facilitate gliding behaviors from the animals (shown in Fig. \ref{fig:species_and_origin}.d for a gliding frog) and calculation of lift and drag from weight balance (see for instance \cite{Tucker1970}).
Wind tunnels enable for precise flow conditions, compared to natural environment or flight arena and facilitate the use of flow visualization techniques, as we will see in the following. \add{Some wind tunnels even emulate hypobaric conditions like the AFAR wind tunnel \cite{Ben-Gida2013,Kirchhefer2013}}. Such experimental conditions are found in 14$\%$ of insect studies for free flight \cite{Ando2002,Johansson2021,LIU1993,Zhang2023,Yokoyama2013,Combes2009,Ortega-Jimenez2013,Evans1997,Crall2017,Ravi2013,Mistick2016,Urca2021}, 17$\%$ of bird research \cite{Ravi2015,Tobalske1999,Tobalske2003,Tobalske2005,Tobalske2007,Tobalske2009,Pennycuick1968,Pennycuick1968a,Ward1999,Ward2002,Tucker1968,Tucker1970,Muijres2012,Dial1997} and 36$\%$ of bat experiments \cite{Cheney2022,Carpenter1985,Johansson2018,Rahman2022,Muijres2011,Muijres2014,Henningsson2018,Riskin2012,Hakansson2015,Hakansson2017,VonBusse2014,VonBusse2012,VonBusse2013,Rummel2019,Hubel2016,Hedenstrom2007,Rayner1986,Iriarte-Diaz2011,Wolf2010}.
\\

For some gliding orders, like gliding squirrels or flying lizards, most studies are made on-site in their environment, with none to limited contact with the animals \cite{Bahlman2013, Koli2011, Krishna2016,Byrnes2008,Byrnes2011,Khandelwal2022}. Gliding experiments are otherwise conducted in flight arenas (Fig. \ref{fig:species_and_origin}.c), which often reproduce natural environment characteristics \cite{Bishop2006,Bishop2008,Paskins2007,Essner2002,Bishop2007,Vanhooydonck2009,Emerson1990,McKnight2020,Young2002}. Fish are most traditionally studied for their swimming but it is interesting to mention that Makiguchi \textit{et al.} performed flight experiments with live flying fishes in an aquarium in 2013 while researching their take-off performance \cite{Makiguchi2013}. As most arboreal frogs tend to parachute more than glide, we only refer here to three studies, that of McCay in 2001 where frogs were placed in front of a wind blower (Fig. \ref{fig:species_and_origin}.d) \cite{McCAY2001} and that of Emerson and Koehl in 1990 and McKnight et al. in 2019, both conducted in a flight arena \cite{Emerson1990,McKnight2020}.

\subsection{Experimental techniques}\label{subsubsec:tech}

With this diversity of experimental conditions comes a wide variety of techniques. The setting of the experiments conditions the techniques that can be implemented and so does the animal species, as we will see hereafter.

Previously in section \ref{sec:flightphysics}, we identified four main observables that constitute the core of animal flight. These variables can be referred to by the acronym TASK, standing for Trajectory, Aerodynamics, Structural deformations and Kinematics. Together, they provide us with a comprehensive view of flight, though they are scarcely observed simultaneously, as noticed in the following.

Across the literature we review here, a number of techniques has been identified for each observable. In particular, seven techniques have been employed for Trajectory reconstruction. Aerodynamics are covered using about 10 experimental techniques, which can be combined as they may contribute to understanding different aspects of the aerodynamics involved, such as flow characteristics or force generation. For Structural deformations, six techniques have been singled out and the same amount for Kinematics.
\\

Flight being at the intersection of Aerodynamics and Locomotion, it is not surprising that most of the experimental techniques stem from both domains. 
\\

Apart from 2D and 3D imaging of trajectories and kinematics that are widely used techniques throughout all domains of physics, engineering and biology, specific techniques were designed for locomotion studies. Among these, Motion Tracking based on Infrared cameras and reflective trackers is often used in flight experiments for Trajectory (part of 3D mapping), Kinematics (part of 3D reconstruction) and even Structural-deformation (3D mapping) measurements, depending on the amount of trackers placed on the animal.
For insects, such trackers may not be necessary, thanks to their high IR-reflectivity \cite{Li2022}.
Structural deformations have also been measured using strain gauges inside bones or muscles and fluoroscope to identify the motion of the skeleton during a movement.
Integrated kinematics, such as flapping frequency, can be measured using accelerometers, which in locomotion research are often used on various places on the animal \cite{Fries2017,White2024} or human \cite{Villeneuve2016}.
Robots mimicking the motion of the animal based on prescribed kinematics are also used in terrestrial locomotion for instance to quantify mechanical efforts, while they can provide insight on aerodynamics in flight studies.

From aerodynamics experiments and more globally from fluid dynamics, many experimental techniques have been adapted to live animal flight. For flow visualization \add{and characterization}, \add{qualitative} smoke stream or \add{quantitative} Particle Image Velocimetry (PIV) have been extensively used with animals, while flow Particle Tracking Velocimetry (PTV) has recently been given more attention as a somehow safer option, \add{with the use of LED lights instead of lasers}. Due to animal welfare, these techniques are not used directly as one would implement them around an airfoil or automobile model. In particular, protections have been designed to prevent any laser-induced injuries to the animals, such as bird laser goggles \cite{Muijres2012,Gutierrez2016} or crossing-triggered laser safety \cite{VanGriethuijsen2006}\add{, like optoisolators \cite{Kirchhefer2013}}.
Like for aerodynamic models, 6-axis Force/Torque sensors have been used in flight experiment, mostly in tethered conditions \cite{Fu2022}, as well as embedded pressure sensors \cite{Usherwood2003}. The latter however require surgery on the animals, which limits their use in most studies and necessitate thorough discussions with ethical committees and local animal welfare bodies.
In addition, Computational Fluid Dynamics (CFD) has been employed to reconstruct flow fields around animals using either static models for gliding flight \cite{LeRoy2021} or moving grid or immersed boundary methods with prescribed kinematics \cite{Liu1998,Yokoyama2013}. A literature review of CFD methods for insect flight can be found in \cite{Engels2022}.
For Structural deformations, the use of laser-shaping technique is also common between standard aerodynamics and animal flight experiment, though the only study we include here, that uses this technique, uses it on wings separated from the animal, which then relates to standard aerodynamics and mechanics studies \cite{HernandezMontoya2023,Ha2015}.
\\

Complementary to these locomotion and aerodynamics techniques, animal flight studies also incorporate experimental methods from meteorology and geography, such as GPS, radar for Trajectory and Kinematics or photogrammetry for Structural deformations.
Methods from other fields of physics are also found such as acoustics and optoelectronic analyzers for Kinematics, PTV for Trajectory reconstruction in flocks, or laser balance using cantilevers \cite{Wolf1993} for Aerodynamics as force measurement. Technology developed for 3D scanners has been further adapted to suit animal flight studies, not only with photogrammetry but also using structured light \cite{Deetjen2017,Deetjen2018}.
\\

In addition to all these techniques that can be found in various domains, two technical innovations have been developed for live animal flight research: the Ornithodolite \cite{Pennycuick1982} and the Aerodynamic Force Platform (AFP) \cite{Deetjen2020}. The former is a device recording the trajectory of a bird using eye-tracking, based on the theodolite device used in land surveys. The AFP relies on control surface integration of fluid forces \cite{Lentink2018,Chin2020}.
\\

As mentioned previously, all these 33 techniques used in live animal flight experiments come with specific adaptations to suit animal welfare and the specifics of the various species. In particular, the same techniques will not be used alike on an insect compared with a bird nor a bat, as they not only differ in size but also in morphology and physiology. The following will thus describe more in details the common implementations of techniques for each powered-flyer group, through a network-based analysis of the literature.
\\

As a guide to the reader, the following three subsections, one per powered-flyer clade, are each organized around a central figure, consisting of a network graph of identified experimental techniques, illustrated with examples from the literature. The graph will be analyzed following the visual hierarchy of its edges: from the largest to the absent, through their density. In particular, every step of analysis will be composed of a visual observation and an interpretation to it. Certain techniques may also be highlighted, to emphasize animal specifics and differences between clades.

The aim of such graphic presentation is to provide an analytical overview of current experimental designs in live animal flight research and identify trends and gaps in technique pairing. Some gaps may be difficult to fill without technological development while others might be critical to our grasp of the physics of flight.

\subsubsection{Insects}\label{subsec:ins}


\begin{figure*}[ht!]
    \centering
    \includegraphics[width=0.85\textwidth]{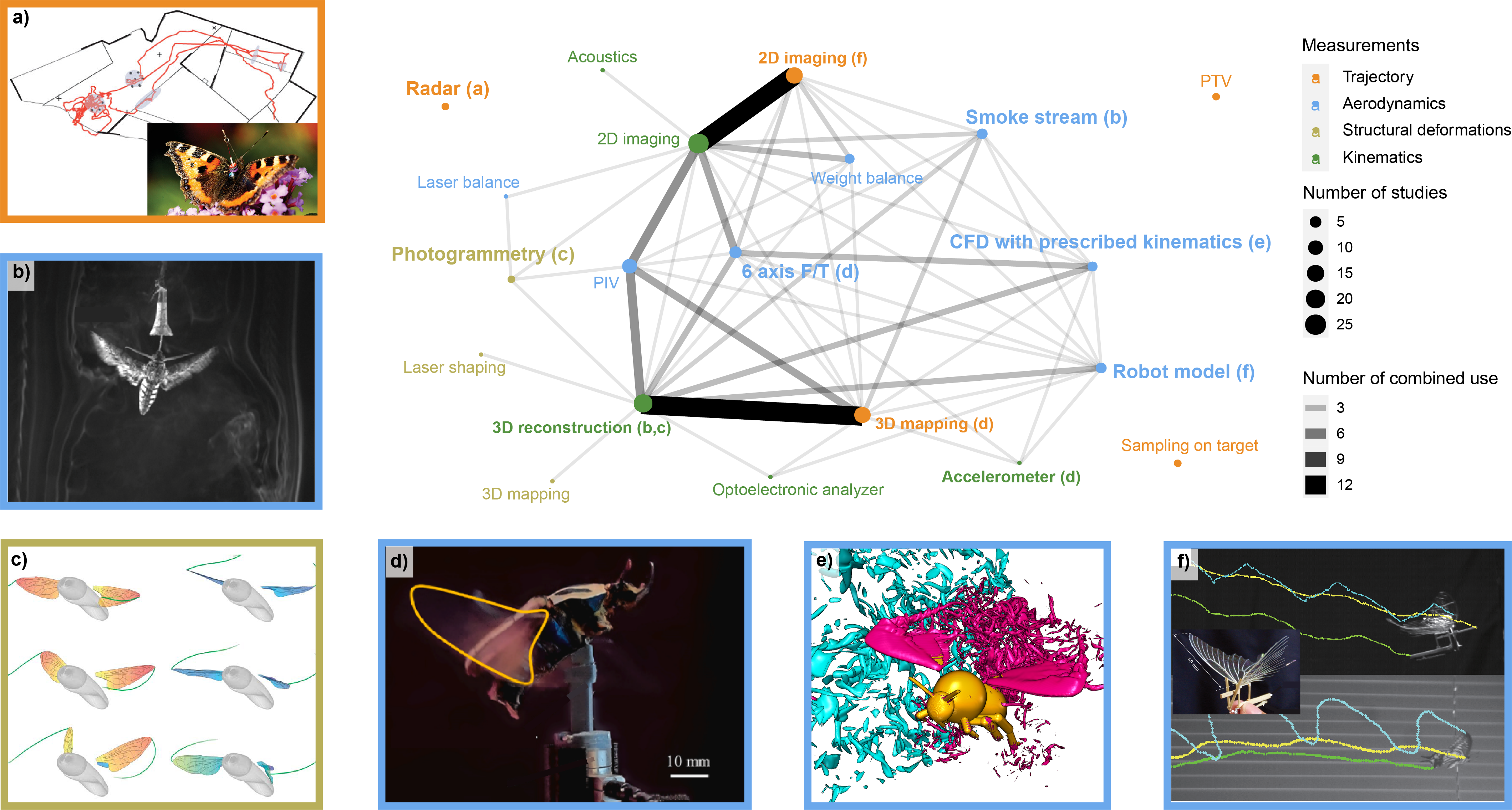}
    \caption{Main experimental techniques used in insect flight research. Center: techniques are linked when used in the same study and their color codes for their observable category defined in Fig. \ref{fig:flight}. Examples are provided for radar tracking (a) \cite{Cant2005}, smoke stream (b) \cite{Ortega-Jimenez2013}, photogrammetry (c) \cite{Walker2009}, 6-axis Force-Torque sensor (d) \cite{Fu2022}, CFD with prescribed kinematics (e) \cite{Engels2016} and the use of robot models (f) \cite{Tanaka2010}. All images are reproduced with permission.}
    \label{fig:expinsects}
\end{figure*}


Insects constitute the largest diversity of flyers but \textit{in vivo} flight experiments are less done with them than birds comparatively (section \ref{sec:landscape}). In terms of implemented techniques though, these experiments are particularly prolific, with 20 techniques listed and 52 pairs of techniques that have been implemented together, as shown in Fig. \ref{fig:expinsects}.
These pairs have not been used in the same amount of studies, with some being extensively reused in the literature.
\\

In particular, two couples of techniques stand out, as both were observed more than 10 times across insect flight studies. These are the coupled use of 2D imaging for Trajectory (T - orange) and Kinematics (K - green) and similarly that of 3D mapping of Trajectory  with 3D reconstruction of Kinematics. Barely connected with one another in one study, they are both T-K combinations, which could be redundant used together. Using 3D reconstruction mostly voids the need for a 2D characterization as it provides more information, unless specific attention to the insect wants to be given that would require high spatial resolution \cite{Combes2009}.
To enhance temporal or spatial resolution, coupling with a second K technique is also possible, like acoustics \cite{Dillon2014} or optoelectronic analyzer \cite{Fry2005}.
\\

Apart from these prominent links, insect experimental techniques are remarkably well-connected, suggesting possible exploratory experiments or species-time constraints. The latter more particularly could relate to the short lifetime of insects that would incite researcher to test and couple many techniques, even redundant, to enhance the reliability of the results despite little reuse possibility of the same individual.
\\

This dense connectivity is especially true for Aerodynamics (A - blue) techniques. Not only are they connected to at least one K technique and mostly also to one T technique, A-A combinations are particularly noticeable. For instance, visualization techniques, such as smoke stream or PIV, are often used conjointly with a quantification of forces, like force sensors or weight balance. Similarly, CFD is often coupled with the use of 6 axis Force/Torque (F/T) sensors. Such combination may be the result of tethered flight experiment for which the force measurements may require a compensation compared to free-flight conditions \cite{Fu2022,Yokoyama2013}.
Likewise, robotic models have been combined with almost all other Aerodynamics techniques, in particular visualization techniques. The joint use can be explained from that small vortical structures observed using smoke stream or PIV may be magnified using scaled-up models, \add{either in air \cite{Ellington1996} or in mineral oil to conserve Reynolds scaling \cite{Fry2005}}. One particular robotic model denotes from the others as it is more of a cyborg than a robot, using motoneuronal control on the flight muscles through a insect-wearable backpack \cite{Fu2022}. This study is also the only insect study reported in this review with accelerometric recording of the flight kinematics. 
Robots otherwise provide the possibility of testing specific mechanisms such as wing rigidity and deformability \cite{Tanaka2010,Johansson2021}.
\\

Other than the two imposing T-K and numerous A-A combinations, little connections are found between Structural deformation (S - ocher) techniques and techniques for any other observable. S techniques are mostly connected to K techniques and few A techniques but their use is overall scarce. More particularly, no coupling T-S has been yet identified in the literature we review here. An explanation to this might be found in the size of insects. Resolving structural deformations at the scale of a millimeter on an insect while tracking decimetric trajectories would require modular fields of view or specific technological development. A pioneering setup towards such combination has recently been developed to enable for high-resolution recording of the insect kinematics over long trajectories by following the insect using a cable-driven automated robot \cite{Pannequin2020}.
\\

The same spatial scale separation could explain the isolation of three T techniques from the rest of the techniques: PTV for the study of swarms in small insects, inducing optical constraints \cite{VanderVaart2019,Sinhuber2017}, radar for observing long-range trajectories \cite{Cant2005,Aralimarad2011} and Sampling-on-target. The use of the latter in flight research is to be correlated with local wind characteristics, using capture and olfactory baits \cite{BRADY1995,Dargas2016}, as the complete trajectory of the animal is not reconstructed. A similar technique is used for arboreal frogs by sampling the horizontal distance covered in a glide \cite{McKnight2020,Emerson1990}.
\\

To summarize, insect flight experiments often combine quite a number of techniques, covering at least 2 TASK observables, Trajectory and Kinematics (T-K), Structural deformation and Kinematics (S-K) or Aerodynamics and Kinematics (A-K). The three-way combination T-A-K has been used several times, using notably PIV for the Aerodynamics. However, no flight study with live insects achieves the conjoint measure of all four TASK observables. Such four-way combination could enable original insights on insect flight, as correlations between observables could be obtained in order to extract causality or simultaneity relations.

\subsubsection{Birds}\label{subsec:bird}

\begin{figure*}[ht!]
    \centering
    \includegraphics[width=0.85\textwidth]{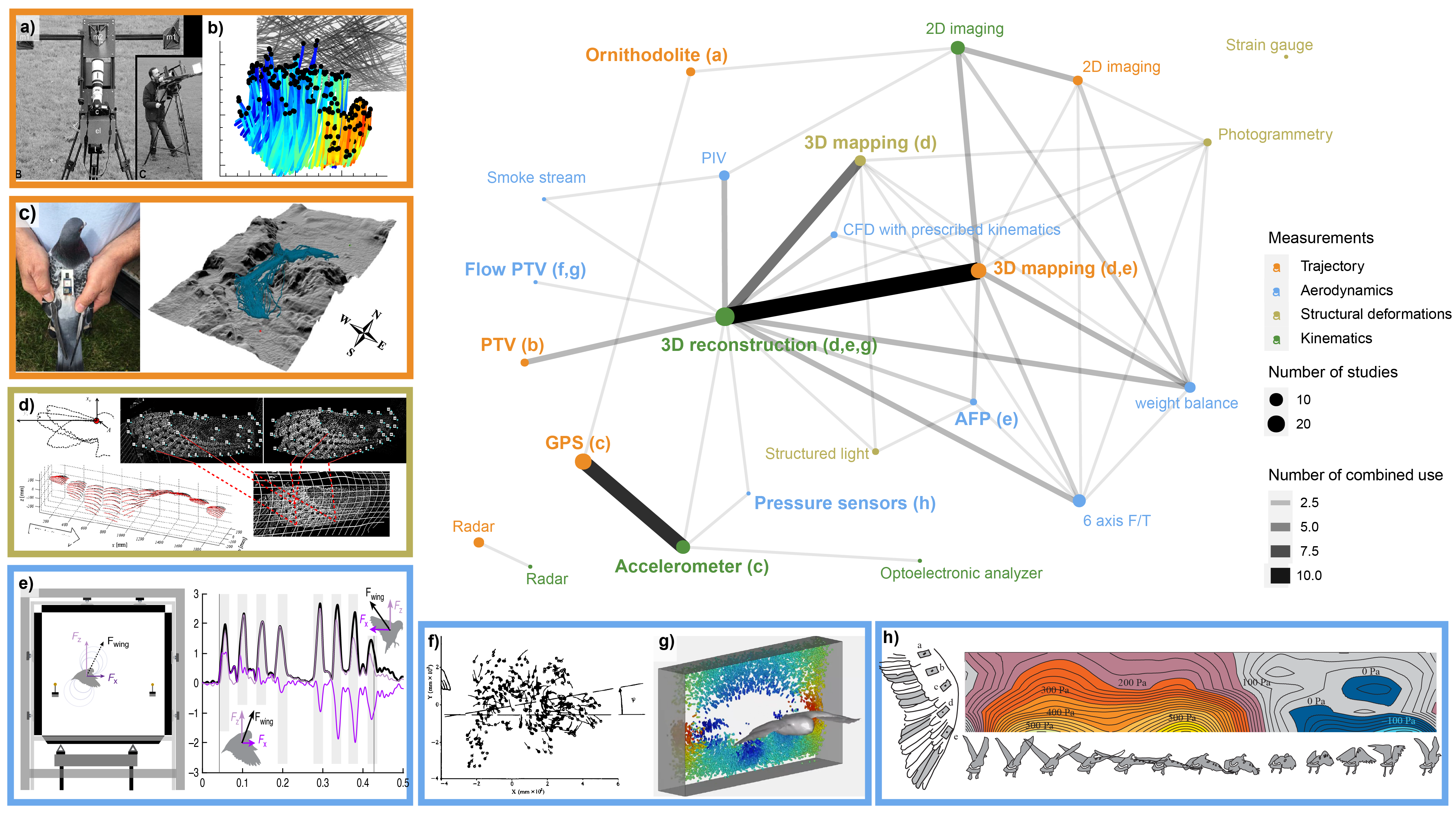}
    \caption{Main experimental techniques used in bird flight research. Center: techniques are linked when used in the same study and their color codes for their observable category defined in Fig. \ref{fig:flight}. Examples of employed techniques are shown for: the ornithodolite (a) \cite{DeMargerie2015}, PTV used with birds as particles (b) \cite{Ling2019}, GPS tracking with accelerometer (c) \cite{Lempidakis2022}, photogrammetry (d) \cite{Wolf2015}, Aerodynamic Force Platform AFP (e) \cite{Chin2019a}, flow PTV (f) \cite{Spedding1984} (g) \cite{Usherwood2020}, and pressure sensors (h) \cite{Usherwood2003}. All images are reproduced with permission.}
    \label{fig:expbirds}
\end{figure*}


Birds are the second largest group of powered flyers but the highest number of live animal flight studies that we report here. It is thus not surprising that they also cover the largest number of implemented experimental techniques totaling 24, as displayed in Fig. \ref{fig:expbirds}. In terms of combined use yet, only 47 pairs have been listed throughout the literature and their organisation widely differs from that of insects.
\\

Bird experiments principally center around two main Trajectory-Kinematics (T-K) axes. While this is similar to insects, the axes themselves and their interactions with the other techniques are much less alike. The larger consists of the combination of 3D reconstruction of the Kinematics (K- green) and 3D mapping of the trajectory (T- orange). The second axis is the combination of GPS (T) and accelerometer (T) in the context of studies in natural environment. The combined use of GPS and accelerometer is extensively used in bird flight as birds are often travelling distances far beyond what a laboratory equipment would allow to track. The miniaturization of GPS tags and accelerometer has also allowed for it to become a less invasive measurement technique and thus expanded its range of application, despite growing quantification of its impact on flight performances \cite{Tomotani2019,Mizrahy-Rewald2023}.
This second axis is however quite isolated from the rest of the techniques due to its unique nature of outdoor experiment. Thus the first axis of 3D T-K measurements can be seen as central in the design of \textit{in vivo} bird flight experiments in laboratory conditions.

These two axes are almost entirely independent from one another, apart from one study using accelerometer as a complement to 3D reconstruction of the kinematics \cite{Usherwood2003}. The use of the accelerometer in that specific study was linked with the use of imbedded differential pressure sensors on the wing in order to compensate for the accelerated component of the differential pressure. Using differential pressure sensors is also a unique characteristic of the study in birds, directly brought from the aerodynamic community as we discussed earlier. 
\\

Despite sharing a dense connectivity between techniques of different observables, contrary to insects, bird experimental techniques for the same TASK observable are scarcely connected. On the contrary, Aerodynamics (A - blue) and Structural deformation (S - ocher) techniques are often coupled with one another as well as with at least Kinematics and even Trajectory reconstruction. Two studies even combines techniques covering all four aspects, one with estimation of the aerodynamic forces using a mass distribution model, as an improved weight balance model \cite{Ros2011} and the other using the AFP with a limited range for the trajectory of about \SI{80}{\centi\meter} \cite{Deetjen2024}.
\\

Like insects however, when Aerodynamics and Structural deformations are measured simultaneously, Trajectory is still mostly discarded, not directly visible in Fig. \ref{fig:expbirds}. Conflicts with Aerodynamics measurements like flow visualization stem from light sources. As flow visualization in air necessitates particularly high powered lighting (LED or laser), T techniques that require light as well are often constrained to the same visualization window. The use of PIV in particular is constrained to a specific position of the animal, trained to maintain it, for safety and for the spatial resolution of the vortices, as we already described in Section \ref{subsubsec:tech}. 

Progress on large scale 3D PTV may enable tracking over larger distance while keeping in mind that optical constraints on the bubble concentration arise as the thickness of the light box increases. Smoke on the contrary is hardly used anymore due to ethical concerns over its harmfulness, compared to low concentrated helium bubbles used in PIV and PTV.
\\

Complementary to flow visualization techniques, Aerodynamics in birds are quantified using techniques very differently from insects. For instance, CFD simulations is mostly used for gliding using a fixed model of the bird \cite{Wang2018,Cheney2020} and robotic models are hardly used, at least not conjointly with live animal experiments \cite{Chang2020}. These two differences stem from the size (intermediate Re) and complexity of birds \cite{Carruthers2010,Harvey2022,Altshuler2015}. Additionally, 6 axis force/torque sensors are used on perches, solely providing insights on forces at take-off and landing. 
This lack of continuous force measurement was recently compensated by the design of the Aerodynamic Force Platform (AFP) \cite{Deetjen2020,Chin2020}. The high technicity of the setup and its implementation yet limit its broader use at the moment.

The AFP is among the five techniques that have been uniquely found in live bird flight experiments (with the exception of one study covering both birds and bats \cite{Ingersoll2018}), together with strain gauge, ornithodolite, structured light and flow PTV. The flow PTV is surprisingly not only unique to birds but also implemented since its early development, \add{where LED panels were flashguns and shake-the-box algorithms stereocomparators} \cite{Spedding1984}.
\\


Three techniques are yet isolated from the rest, strain gauges (S) and radar measurements (T-K).

Strain gauges can be placed inside feathers, muscles or bones \cite{Corning1998,Williamson2001}. Kinematics are usually not recorded but estimated from the gauge recording. Such gauges otherwise serve for internal actuation quantification.

Conversely radar studies are isolated due to their outdoor use, for migration and understanding wind influence on large-scale trajectories. Mostly providing insight on the flight speed and trajectory of numerous animals, it may even be used for Kinematics in specific species like swifts \textit{Apus apus} \cite{Backman2001}.
\\

Additional outdoor T techniques consist of the ornithodolite, introduced in \ref{subsubsec:tech} \cite{Pennycuick1982} and coupled with GPS tracking \cite{MIZUGUCHI2017} and 2D imaging of the kinematics \cite{Pennycuick2001}, and PTV tracking of the animals \cite{Ling2018,Ling2019,Shelton2014}. For the latter, depending on the field of view and the number of individuals, Kinematics might be reconstructed from the same images, however only coarsely. The advantage of the PTV however it the simultaneous tracking of large numbers of animals, which suits bird flocks, just like insect swarms.
Overall, for both birds and insects, outdoor techniques are scarcely used combined with other techniques, due to their inherent large scale characteristics.
\\

To conclude, bird experiments tend to combine between two and three TASK observables, yet with only one technique per observable, leaving little redundancy. This provides a rather efficient data production, which is yet limited by its precision and necessitates thorough calibrations prior to the live animal runs. This may however be explained by the fact that every new technique introduced could require additional training and extend significantly the duration of any experiment.


\begin{figure*}[ht!]
    \centering
    \includegraphics[width=0.9\textwidth]{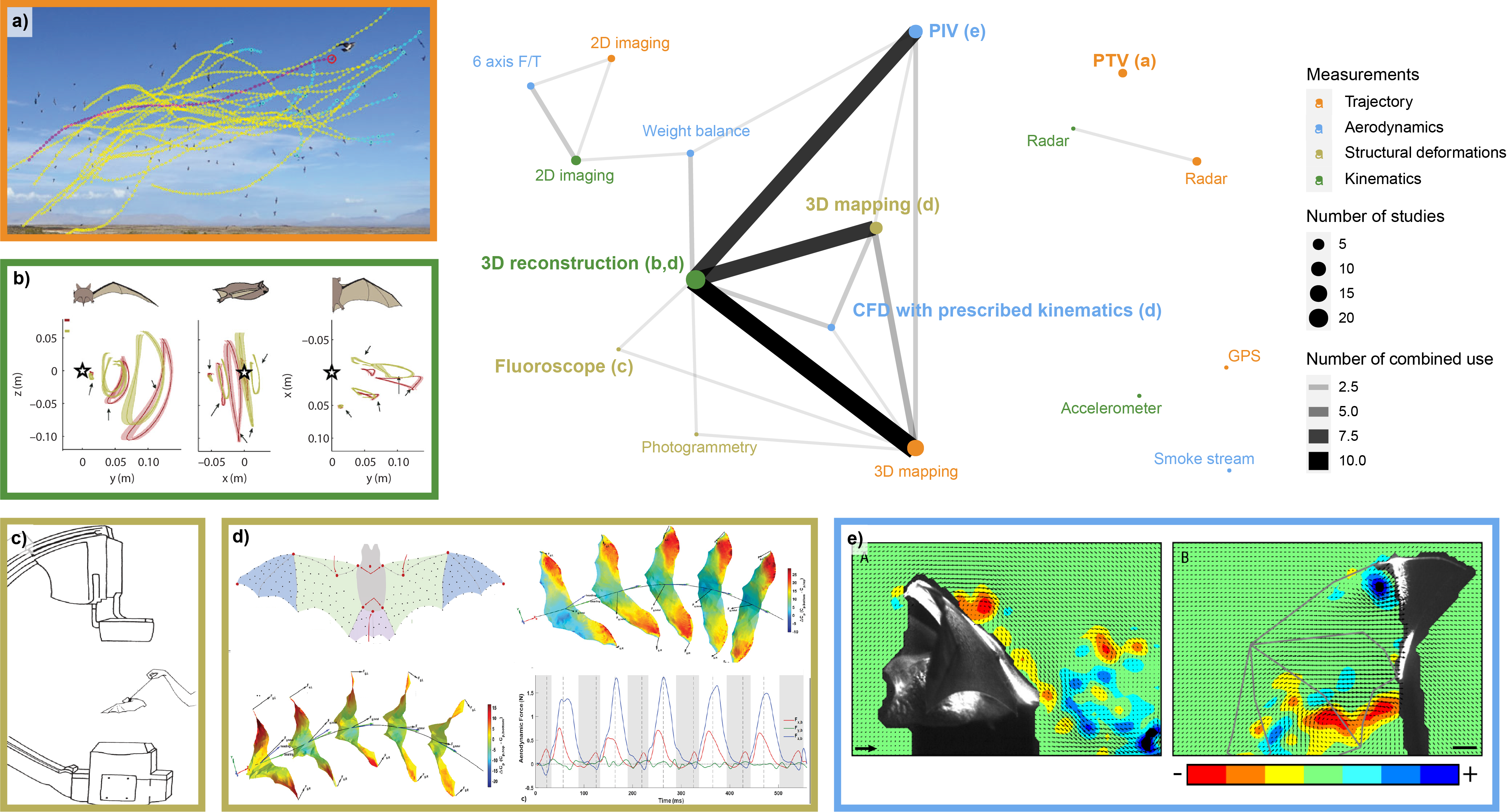}
    \caption{Main experimental techniques used in bat flight research. Center: techniques are linked when used in the same study and their color codes for their observable category defined in Fig. \ref{fig:flight}. Examples of techniques are shown for: PTV tracking of bats (a) \cite{Brighton2022}, 3D reconstruction of body kinematics (b) \cite{Hubel2016}, fluoroscope (c) \cite{Panyutina2013}, 3D mapping of the deformation with CFD simulations for aerodynamics (d) \cite{Windes2020}, PIV measurements around the animal (e) \cite{Muijres2014}. All images are reproduced with permission.}
    \label{fig:expbats}
\end{figure*}


\subsubsection{Bats}\label{subsec:bat}

As fellow vertebrates, bats present similar constraints as birds do. Tethered flight experiments are mostly prescribed and training needs limit the combined use of techniques. Together with a smaller number of studies, the diversity of experimental techniques is narrower than for birds and insects, with only 17 techniques and 20 pairings, as shown in Fig. \ref{fig:expbats}.
\\

Unlike birds and insects, bat flight experiments organize themselves along three combinations, anchored by the 3D reconstruction of the flight Kinematics (K - green). Each combination links kinematics to another observable: PIV for Aerodynamics (A - blue), 3D mapping for Trajectory (T - orange) and 3D mapping for Structural deformation (S - ocher). Each is used in about 10 studies, for a total of 50 reported publications. In comparison, other combinations are hardly found in more than 2 to 5 studies. 
\\

Conjointly to this key triad, a noticeable feature of bat experiments is the reduced complexity of the graph, with little connections between the different techniques as we mentioned earlier. In particular, apart from the A-A combination between PIV and weight balance, no connection is found between two techniques of the same observable, avoiding redundancy to an even greater length than in bird flight. 
\\

Compared to insects, 2D imaging (T-K) and 6 axis F/T sensors are rather estranged from the rest of the techniques, only connected to the main techniques through weight balance, though it might be just a bias due to the rare use of 2D imaging. 3D reconstruction and mapping was indeed introduced notably early in bat flight research \cite{Aldridge1986,Hughes1991}.
\\

In addition to that, four techniques are fully isolated (smoke stream for Aerodynamics, accelerometer for Kinematics, PTV and GPS for Trajectory) and a group of two other techniques (radar for Trajectory and Kinematics) is also separated from the core techniques. 
Like for birds, radar has been used for tracking large-scale motion in bats, while contrastingly GPS has scarcely been used. This might be explained by the lifestyle of bats, living in caves, as GPS signals have limited penetration underground. Bat behaviors may yet allow for unique opportunities for technical development. For instance, a combination of acoustics and LiDAR techniques has recently been proposed to reconstruct trajectories of bats for behavioral studies \cite{Hermans2023}. This novel technique is, in particular, enabled by the echolocation behavior of the bats, almost constantly feeding microphone recording for backtracking.
\\

Other similarities with birds are found in the reduced use of smoke stream and force sensors. The latter is only used either for take-off characterization \cite{Schutt1997} or with prepared specimens \cite{Carpenter1985}. With less details on the wings than birds, photogrammetry is not a common technique in bats, while markers for 3D mapping are easier to place on the smooth membrane and bony parts of the wings, compared to feathers.
\\

Interestingly, a S technique at the moment uniquely used in bats is the fluoroscope. A possibility to this uniqueness is the difficulty of high resolution of anatomical movement using fluoroscopy. In particular, resolving bone motion from X-ray imaging is constrained by the bone density. While birds have hollow pneumatized bones, bats have flexible yet dense mammalian bones which present higher contrast on the fluoroscope imaging. Due to their exoskeleton and smaller size, insects are especially difficult to image with this technique. Micro-CT scanning could solve this issue but constrains air flow and flapping motion, limiting the impact of any there-obtained insight. 
\\

To conclude, bat experiments present close traits to bird experiments in the study of flight with yet a very different implementation. This might be linked to the reduced number of studies but also to the fact that bat colonies are hosted in only few universities across the world, thus narrowing technique variety.

Yet despite the limited range of techniques compared to insects and birds, bat flight studies have been thorough in the characterization of flight across species and tend towards the observation of a "universality" of flow phenomena in bat flight \cite{Hedenstrom2015}, which is yet to be truly observed in the other flyers, which also present a higher diversity (Fig. \ref{fig:species_and_origin}).

\subsection{Towards future designs}\label{subsec:future}
As we observed, the large variety of \textit{in vivo} flight experiments still struggle to cover the immense diversity of animal flight.
Many directions of research are yet to be explored in the physical understanding of animal flight. 
\\

A possible direction is to extend the knowledge of the different flights of a specific order, like dragonflies, and map the flight characteristics in a multitude of parameters, both physical (\textit{e.g.} flow properties) and biological (\textit{e.g.} wing shapes). Such direction, close to that currently observed for bats, aims at identifying key mechanisms in flight and often looks for reduced order models, with limited control parameters. This approach has recently emerged for Morpho butterflies \cite{LeRoy2021} and hovering insects \cite{Jung2023}.

Another direction could be to homogenize \textit{in vivo} flight experimental knowledge across different orders. For instance, it could be about realizing the same experiment for two widely different animals, bird and bat or reptile and gliding mammal for instance, with techniques as similar as can be despite physiological differences. This has already been initiated for birds and bats regarding wake structures and aerodynamic forces \cite{Muijres2012a,Ingersoll2018} and for all three powered-flyer clades for take-off load capacity and lift production \cite{Marden1987}. This homogenization strives particularly to find universal processes, often to be applied to engineering concepts.

Based on the gaps identified in our network analysis of the literature, a third direction could be to design reference experiments that cover all four TASK observables at the same time, linking expertise across fields. In particular, these experiments could lead to the establishment of a broad database of flight kinematics for any species, that one could use for numerical simulations, bio-inspired engineering or large-scale data-driven modelling. Such design can be found in its early stages in birds \cite{Ros2011,Deetjen2024} and bats \cite{Rahman2022}, but has yet to be standardized across clades.
\\

Yet, all these directions share a common trait: the animal and the comprehension of its flight are the central object of interest.
In the following section, we propose a fourth direction of research, that centers around the physical phenomena involved in flight, leveraging the animal natural adaptability.


\section{\label{sec:AL}Rethinking environmental parameters}

Throughout the literature, the use of live animals in the experiment is largely driven by the question: how does it fly? 
Such question resonates with the ontogeny and causation questions from Tinbergen in ethology \cite{Tinbergen1963}.
In particular, when investigating causation, the introduction of environmental constraints is often motivated by questioning the limits of flight abilities and looking at how the animal has adjusted itself to its new flight environment.
This specific question is tackled in the literature along two main approaches.
\\

First, we find a bottom-up approach, closely related to engineering and physical sciences, in which the complexity of the problem is increased at each step of the understanding. For instance, to understand the aerodynamics of a particular insect species, a first step is to place a prepared specimen in a wind tunnel before placing a live tethered one. Before even switching to free flight, an elaboration on the flow structure can be performed, through the introduction of gusts and vortical structures, such as vortex streets, all the way to fully developed turbulence.

The second approach resonates more with biology, as a top-down approach. The system is first viewed as a whole before removing layers of complexity. An example of such approach is found in the study of gliding mammals that begin with observations in their natural environment, followed by flight arena experiments. In a similar even more advanced top-down development, the understanding of insect flight was brought from initially observing the formation of vortex rings on flying hawkmoths and then reproducing the motion on a scaled-up robot to further refine the comprehended phenomenon, reported in Fig. \ref{fig:lit}.e-f \cite{Ellington1996}.

These two approaches are intrinsically connected as the end product of one is the starting point of the other and vice-versa. For instance, while flies were decomposed into rotating flat plates \cite{Sane2001,Lentink2009a} from earlier tethered flight experiments \cite{Lehmann1997,Lehmann1998}, recent progress on CT scanning enabled a better resolution of the structure of fly wings for implementation in CFD simulations \cite{Engels2020}.
\\

Another way at looking at animal flight, which we will further develop throughout this section, is to ask the question: why can it fly?
This question can connect to comparative and evolutionary biology as well as physics in different contexts. From a biological point of view, ``why" relates to Tinbergen's ultimate questions, phylogeny and function \cite{Tinbergen1963}.  These correspond to the `time and space' variations of the animals themselves. On the contrary, the physics perspective investigates the parameter space surrounding the animal. It can be understood as what physical mechanisms make it possible for the animal to fly this way.
The modifications of the environmental constraints are then driven by their influence on the animal, rather than the limits they impose.
\\

The question ``why" assumes that the animal does fly and does not require a full knowledge of the ``how". Similarly, we can answer the ``how" without questioning the ``why". However, it is undeniable that these two questions are coupled with one another and any experiment designed to answer one will provide insights on the other as well. For instance, the same study that explained how hawkmoths fly using the leading-edge vortex also covered a part of the why, providing novel physical understanding of lift production \cite{Ellington1996}. Reciprocally, the question of why geese fly in V-formation led to the later comprehension of the specific spacing between birds in such formation, thus answering part of the ``how" \cite{Lissaman1970,Portugal2014}.
\\


\begin{figure*}[ht!]
    \centering
    \includegraphics[width=0.90\textwidth]{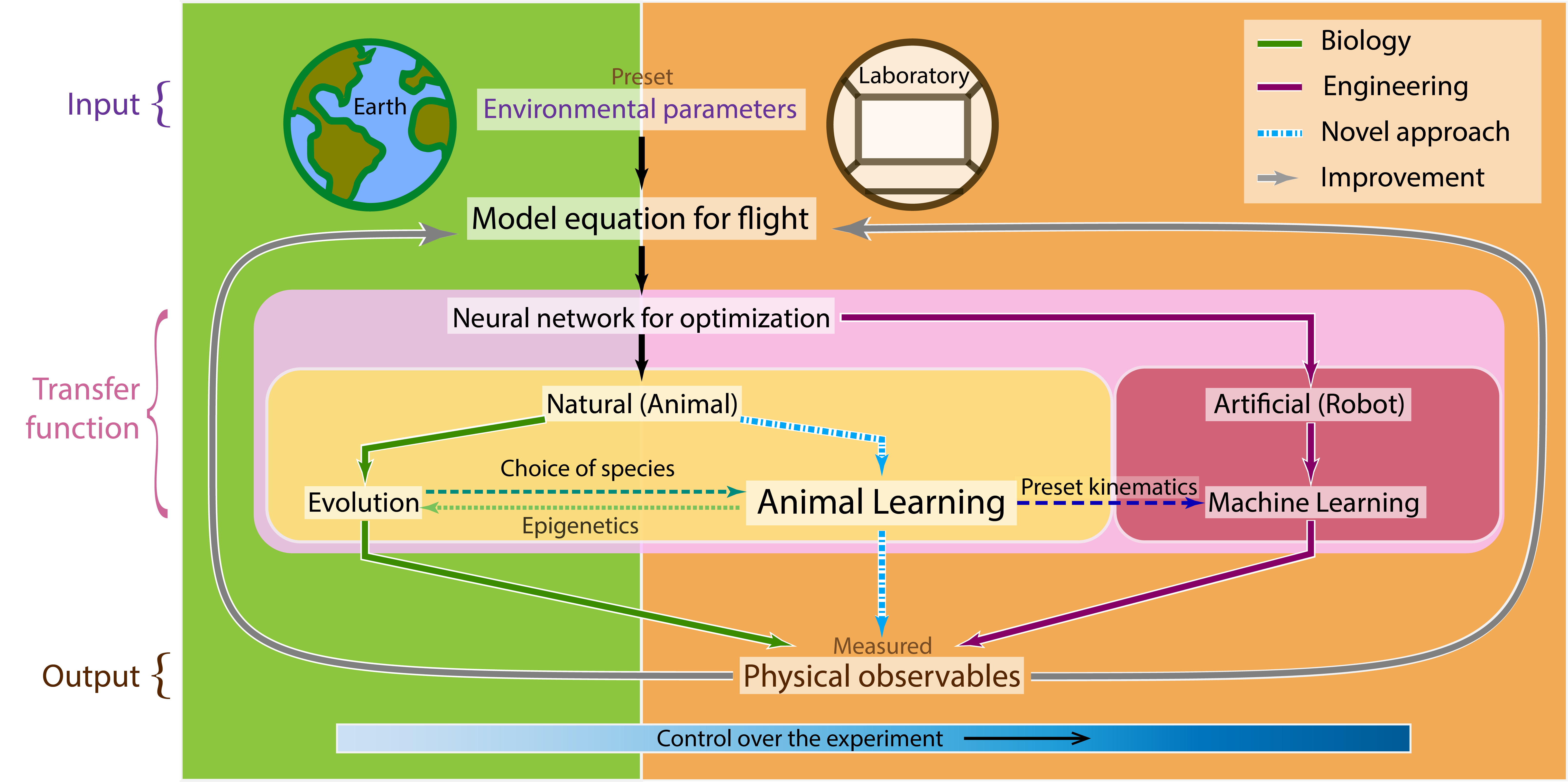}
    \caption{Schematic representation of the concept of Animal Learning, where animals become probes to study physical phenomena. Within the context of developing a physical model for flight, Animal Learning (dashed blue) is compared with widespread biological (dark green) and engineering (purple) pathways, which rely respectively on evolution and machine learning. Environmental parameters cover for instance: air density $\rho$, atmospheric pressure $P$, temperature $T$, viscosity $\nu$, humidity $RH$, gravity $g$, turbulence rate $\varphi$... These parameters are only measured variables in the Earth framework (left - green) while they can be controlled in the lab framework (right - orange).}
    \label{fig:concept}
\end{figure*}


As the nuance between the two questions is especially narrow, let us imagine a hypothetical experiment centered around hovering flight in kestrels. 

If the question we ask is the ``how", the focus of the experiment will be on the animal and more particularly the characterization of its flying motion in conditions where we know this type of flight happens. In addition, some assumptions might be made on the force balance based on the stability of the flight or on the flow characteristics that enable this flight, founded on observations. Such approach has recently been pursued with a thorough characterization of body kinematics of hovering nankeen kestrels \cite{MartinezGroves-Raines2024}.

Yet, if the central question is now the ``why", the core interest will be on the flow surrounding the animal and the experimental conditions. For instance, as hovering behavior frequency is observed to be strongly dependent of the wind speed for both kestrels \cite{Videler1983} and ospreys \cite{Strandberg2006}, a question of the ``why" would be to correlate kinematics to flow velocity and the proportion of gusts in the incoming wind. In these conditions, the details of the kinematics would be more of a by-product of the investigation than the main interest, as the relevant parameter is their evolution with the experimental flow constraints.
\\

This can be even further accentuated into an original research paradigm if we no longer consider the animal as the object of interest but rather as a probe to physical phenomena involved in flight.

This novel paradigm, which we propose here, is centered around the concept of Animal Learning, where the natural neural network of the animals is harnessed to explore flow physics through flight response.
\\

A schematic representation of the use of Animal Learning in animal flight is shown in Fig. \ref{fig:concept}. From a theoretical point of view, flight is the observed output of an animal action on its environment, such that the environmental parameters can be seen as inputs and the animal as a transfer function:
\begin{equation}\label{eq:h(x)}
    \mathrm{Flight} = \mathrm{Animal}(\mathrm{Environment})
\end{equation}

The most obvious object of interest is then the characterisation of the transfer function, \textit{i.e.} the animal.
To build and refine a model equation for flight, one can use a neural network to explore the input parameter space for finding optima in the model. Optima might be found in various variables, such as energy, trajectory, kinematics, lift/drag ratio, \cite{KleinHeerenbrink2022,Harvey2024}. Which optima is relevant depends on the desired model.

A recent way of exploring this space is the use of Artificial Intelligence. For instance, this could be achieved by designing a robot and prescribing kinematics across a wide range of parameters, such as flapping frequency and amplitude or wing-morphing trajectories \cite{Hoey2010,Karasek2018,Chang2020}.
Machine Learning or genetic algorithms are then used to find local optima within the parameter space through extensive scanning \cite{Melis2024,Busch2022}. After optimization, the model is refined based on the measured physical variables. This approach relates closely to engineering (purple pathway in Fig. \ref{fig:concept}).

Another possibility is the use of a natural neural network, thus an animal itself. The most common approach is the consideration of the animal as a species and looking at its evolution from a biological point of view (green pathway in Fig. \ref{fig:concept}). The then used neural network has been shaped and trained over generations of animals and depending on the selectivity of flight on its survival, reached a certain optimum. Placing such animal in its natural environment provides insights on flight development and specific techniques like maneuvering for instance. Looking at species differences also strengthens any model of flight as various morphologies enable for different aerodynamic mechanisms \cite{LeRoy2021}.
\\

In-between these two paths to model refinement, we propose a novel approach using the animal neural network in controlled conditions as a probe, through its own learning capabilities as an individual (dashed blue path in Fig. \ref{fig:concept}).

In particular, if we come back to Eq. \ref{eq:h(x)}, we propose to rethink this equation with a second transfer function hidden in all flow-animal interactions at play, which would act as a convolution of animal action and flow retro-action on the animal motion. Eq. \ref{eq:h(x)} then becomes: 
\begin{equation}\label{eq:h2(x)}
    \mathrm{Flight} = \mathrm{Flow}*\mathrm{Animal}(\mathrm{Environment})
\end{equation}

As physicists, our aim is to proceed to the deconvolution of both transfer functions, and in particular isolate the flow component from the animal response. To this end, it is possible to repurpose environmental parameters towards a statistical estimation of the animal. The natural adaptability of the animal is then both a curse and a blessing.
It is a curse because it means that its response to stimuli will be modified throughout the experiment, biasing the statistics. On the other hand, it is also a blessing as it allows for finding local optima and the time evolution of the response to the novel environment can also provide interesting insights on the animal.

The parameter controlled in the experiment then becomes an important object of interest in the system as is the animal. The core attention still lies in the resultant flight but the focus is on variations of the animal transfer function.

For instance, a question we can address using this concept is the difference in influence between atmospheric pressure and air density on vortex formation. Taking an insect which presents a vortex-dominated flight like a bumblebee, we can place this insect in various laboratory environments to investigate this difference. One environment would be a pressure-regulated chamber allowing for hyperbaric or hypobaric conditions \cite{Dillon2014}. A second environment would be a HeliOx chamber where air density is varied using a gas mixture of helium and oxygen \cite{Altshuler2003,Dudley1995}. Matching air density in both environments and looking at the variations of flight kinematics between the two environment for the same insect may provide interesting insights on the vortex dependence on atmospheric pressure, which is not yet modelled by Navier-Stokes equations.

Note that the Animal Learning approach also gives insight on the animal biomechanics as well and connections between the three paths are also important. For example, any kinematics obtained through Animal Learning may be used as initial conditions to a machine-learning process and evolution provides a selection of species which are more likely to enable the Animal Learning success. A lesser-known connection is the possibility of extending Animal Learning towards experimental evolution thanks to epigenetics, as individuals may pass on experience to their offspring outside of genes. This was in particular observed in flight experiments in microgravity using quails (\textit{Coturnix japonica}) \cite{Bilcik1996}. 
\\

Hence, from a physical point of view, Animal Learning can be seen as harnessing millions of years of evolution and training to improve our understanding of flow phenomena. Such harness has been recently made possible through the extensive technical progress described throughout this article and comes to complement already existing animal flight research from a different angle, at the intersection of many disciplines. Maybe such paradigm could also be reversed in the future where we use flying animals to directly probe atmospheric characteristics, as already seen for birds \cite{Laurent2021,Lempidakis2022,Jetz2022}, in a data-driven methodology.



\bibliography{biblio_february2025}

\end{document}